\documentclass[aps,prl,twocolumn,amsmath,amssymb,nofootinbib,superscriptaddress,floatfix,reprint,longbibliography]{revtex4-1}
\usepackage[dvips]{graphicx}
\usepackage{latexsym}
\usepackage{amsmath}
\usepackage{amsfonts}
\usepackage{amssymb}
\usepackage{color}
\usepackage{txfonts}
\usepackage{float}
\usepackage{url}
\usepackage[colorlinks=true, urlcolor=blue, linkcolor=blue, citecolor=blue, pdftex]{hyperref}
\usepackage{ulem}
\usepackage{graphicx}
\usepackage{extpfeil}
\usepackage{subfigure}
\usepackage{physics}
\DeclareUnicodeCharacter{2212}{-}

\begin{document}
	\newcommand{\fig}[2]{\includegraphics[width=#1]{#2}}
	\newcommand{\la}{{\langle}}
	\newcommand{\ra}{{\rangle}}
	\newcommand{\dg}{{\dagger}}
	\newcommand{\upa}{{\uparrow}}
	\newcommand{\dna}{{\downarrow}}
	\newcommand{\ab}{{\alpha\beta}}
	\newcommand{\ias}{{i\alpha\sigma}}
	\newcommand{\ibs}{{i\beta\sigma}}
	\newcommand{\hH}{\hat{H}}
	\newcommand{\hn}{\hat{n}}
	\newcommand{\hc}{{\hat{\chi}}}
	\newcommand{\hU}{{\hat{U}}}
	\newcommand{\hV}{{\hat{V}}}
	\newcommand{\br}{{\bf r}}
	\newcommand{\bk}{{{\bf k}}}
	\newcommand{\bq}{{{\bf q}}}
	\def\gsim{~\rlap{$>$}{\lower 1.0ex\hbox{$\sim$}}}
	\setlength{\unitlength}{1mm}
	\newcommand{{\vhf}}{$\chi^\text{v}_f$}
	\newcommand{{\vhd}}{$\chi^\text{v}_d$}
	\newcommand{{\vpd}}{$\Delta^\text{v}_d$}
	\newcommand{{\ved}}{$\epsilon^\text{v}_d$}
	\newcommand{{\vved}}{$\varepsilon^\text{v}_d$}
	\newcommand{\pprl}{Phys. Rev. Lett. \ }
	\newcommand{\pprb}{Phys. Rev. {B}}

\title {The Mottness and the Anderson localization in bilayer nickelate La$_3$Ni$_2$O$_{7-\delta}$}

\author{Yuxin Wang}
\affiliation{Beijing National Laboratory for Condensed Matter Physics and Institute of Physics,
	Chinese Academy of Sciences, Beijing 100190, China}
\affiliation{School of Physical Sciences, University of Chinese Academy of Sciences, Beijing 100190, China}

\author{Ziyan Chen}
\affiliation{Beijing National Laboratory for Condensed Matter Physics and Institute of Physics,
	Chinese Academy of Sciences, Beijing 100190, China}
\affiliation{School of Physical Sciences, University of Chinese Academy of Sciences, Beijing 100190, China}

\author{Yi Zhang}
\email{zhangyi821@shu.edu.cn}
\affiliation{Department of Physics and Institute for Quantum Science and Technology, Shanghai University, Shanghai 200444, China}
\affiliation{Shanghai Key Laboratory of High Temperature Superconductors and International Center of Quantum and Molecular Structures, Shanghai University, Shanghai 200444, China}     

\author{Kun Jiang}
\email{jiangkun@iphy.ac.cn}
\affiliation{Beijing National Laboratory for Condensed Matter Physics and Institute of Physics,
	Chinese Academy of Sciences, Beijing 100190, China}
\affiliation{School of Physical Sciences, University of Chinese Academy of Sciences, Beijing 100190, China}

\author{Jiangping Hu}
\email{jphu@iphy.ac.cn}
\affiliation{Beijing National Laboratory for Condensed Matter Physics and Institute of Physics,
	Chinese Academy of Sciences, Beijing 100190, China}
\affiliation{Kavli Institute of Theoretical Sciences, University of Chinese Academy of Sciences,
	Beijing, 100190, China}
 \affiliation{New Cornerstone Science Laboratory, 
	Beijing, 100190, China}

\date{\today}

\begin{abstract}
The oxygen content plays a pivotal role in determining the electronic and superconducting properties of the recently discovered La$_3$Ni$_2$O$_{7-\delta}$ superconductors. In this work, we investigate the impact of oxygen vacancies on the insulating behavior of La$_3$Ni$_2$O$_{7-\delta}$ across the doping range $\delta = 0$ to $0.5$. At $\delta = 0.5$, we construct a bilayer two-orbital Hubbard model to describe the system. Using dynamical mean-field theory, we demonstrate that the model captures the characteristics of a bilayer Mott insulator. To explore the effects of disorder within the range $\delta = 0$ to $0.5$, we treat the system as a mixture of metallic and Mott insulating phases. By applying the dynamical cluster approximation and the typical medium dynamical cluster approximation, we identify an Anderson localization transition at a critical doping of $\delta \sim 0.2$ through the geometric average of the local density of states. This Anderson localization transition is the key reason for the suppression of superconductivity in La$_3$Ni$_2$O$_{7-\delta}$. These results provide a quantitative explanation of recent experimental observations and highlight the critical influence of oxygen content on the physical properties of La$_3$Ni$_2$O$_{7-\delta}$.
\end{abstract}
\maketitle

\begin{figure}
	\begin{center}
		\fig{3.4in}{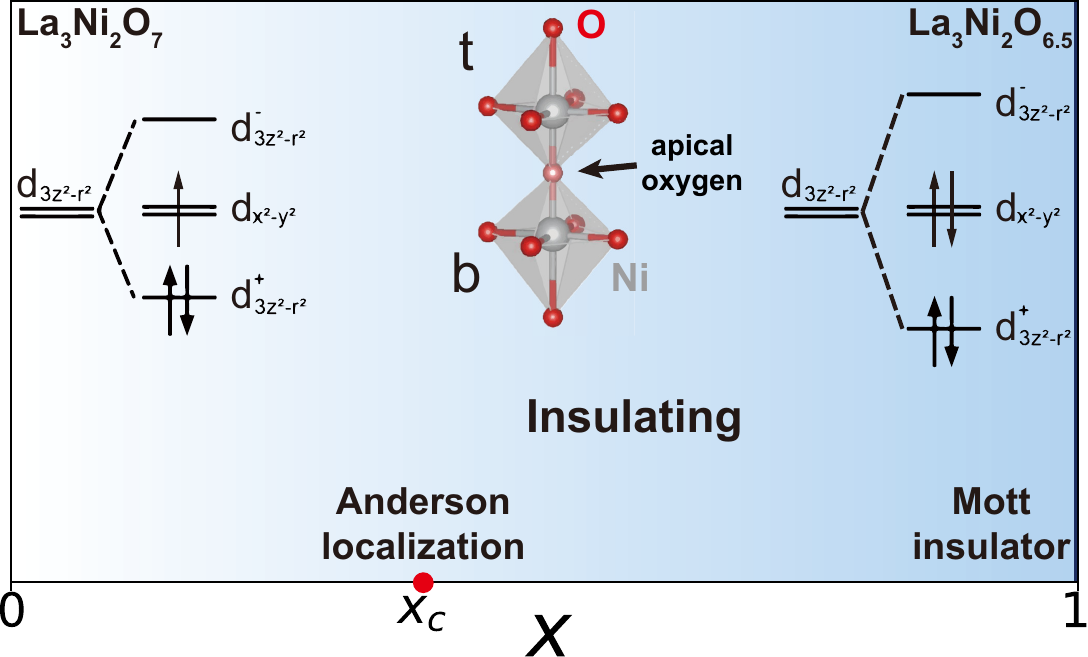}
		\caption{The phase diagram of La$_3$Ni$_2$O$_{7-\delta}$ in the region of $\delta=0-0.5$. The variable $x$ represents the additional filling in the $e_g$ orbitals of Ni introduced by the creation of O vacancies in La$_3$Ni$_2$O$_{7-\delta}$, with $x=2\delta$ due to the chemical valence of O is O$^{2-}$. The crystal field splitting is also shown here and the $d_{x^2-y^2}$ orbital is exactly half-filled at $x=1$. We also plot the enlarged bilayer Ni-O octahedral structure for large $x$, in which the corner shared apical pink ball can be regarded as 0.5 oxygen for $x=1$ in VCA. The content of apical oxygen vacancies has a significant impact on superconductivity. The $t$/$b$ indicates the layer index.  }\label{fig1}
	\end{center}
\end{figure}

The recent discovery of high-temperature superconductivity in bilayer nickelate La$_3$Ni$_2$O$_7$ under pressure (HP) greatly spurs the extensive research efforts in nicklelate superconductors \cite{meng_wang,chengjg_crystal,chengjg,yuanhq,chengjg_poly,sunll,ni_review}. Many theoretical works are dedicated to understanding the origin of the high-temperature superconductivity and the corresponding pairing properties \cite{yaodx,dagotto1,wangqh,Kuroki,guyh,zhanggm,werner,yangf,wucj,dagotto2,yangyf2,ryee2023critical,kun_cpl,PhysRevB.108.L201121,PhysRevB.109.115114}. Many subsequent experimental studies have further enhanced our understanding and led new directions for theoretical work \cite{zhouxj,meng_wang2,rixs,nmr,musr,musr2}.
Now, it is well established that the bilayer structure of La$_3$Ni$_2$O$_7$ is central to its superconductivity \cite{meng_wang,ni_review}. The two NiO$_2$ planes must be regarded as a strongly coupled bilayer unit. This stands in clear contrast to bilayer cuprates, where the interlayer coupling is relatively weak. As shown in Fig.\ref{fig1}, the interlayer coupling in La$_3$Ni$_2$O$_7$ is mediated by the apical oxygen. Therefore, understanding the role of apical oxygen is essential for elucidating its electronic properties and superconducting mechanism.

Experimentally, oxygen vacancies in  La$_3$Ni$_2$O$_{7-\delta}$ have been consistently observed to reside at the apical oxygen sites \cite{chenzhen-TEM}. Moreover, recent studies indicate that superconductivity in  La$_3$Ni$_2$O$_{7-\delta}$ emerges only for ($\delta \leq 0.1 $, electron-doped regime) \cite{ueki2024phase,sunll}, while the compound exhibits insulating behavior over a broad range of higher vacancy concentrations ($\delta=0.16-0.5$) \cite{ueki2024phase,cava_3265,327_structure1,327_structure2}. These results demonstrate that apical oxygen vacancies strongly alter the electronic structure of La$_3$Ni$_2$O$_{7-\delta}$ and suppress superconductivity. Understanding this effect is therefore a central question that we aim to address.

In this work, we carry out a systematic study of the La$_3$Ni$_2$O$_{7-\delta}$ within the range of $\delta=0\sim0.5$, aiming to clarify the insulating behavior in this region and the role of apical oxygen vacancy. The main conclusions are summarized in Fig. \ref{fig1} and Ref. \cite{ueki2024phase}. 
Given that the chemical valence of O is O$^{2-}$, we follow the convention using electron doping $x=2\delta$ in cuprates. 
The $x=0$ phase is La$_3$Ni$_2$O$_7$, which is metallic and superconducting at high pressure.
The La$_3$Ni$_2$O$_{6.5}$ at $x=1$ is a Mott insulator \cite{ueki2024phase,cava_3265,Pickett}.
Among the doping evolution, the system becomes Anderson insulating after the critical doping $x_c$ around 0.4.


Much like the extensively studied La$_3$Ni$_2$O$_7$, the most significant part of the crystal structure of La$_3$Ni$_2$O$_{7-\delta}$ is also the bilayer Ni-O octahedral block. However, the difference lies in the fact that the two octahedrons do not tilt even at ambient pressure for large $\delta$, resulting in only two Ni atoms per unit cell, which we can label them as $t$ (top layer) and $b$ (bottom layer), as shown in the inset of Fig. \ref{fig1}. The crystal structure of $\delta=0.16$ retains the orthogonal $Amam$ space group symmetry, similar to $\delta=0$ \cite{ueki2024phase}. However, for $\delta=0.5$, the tetragonal $I4/mmm$ space group is adopted, with the lattice parameters $a=3.874\AA$ and $c=20.075\AA$ \cite{ueki2024phase}. There should be an average of 0.5 oxygen vacancies in each unit cell. Previous studies have shown that oxygen vacancies are primarily located at the corner-shared apical oxygen site between the two Ni-O octahedra (marked in pink in Fig. \ref{fig1}) \cite{dong2024visualization,ueki2024phase}. Using the virtual crystal approximation (VCA), we can model this vacancy by setting 0.5 oxygen atoms at this site in each unit cell. From this, it is evident that there are symmetry operations can map $t$-Ni into $b$-Ni, or vice versa, rendering the two Ni atoms equivalent.

The octahedron crystal field in each layer splits the five degenerate Ni 3$d$ orbitals into two groups: $e_g$ and $t_{2g}$. For $x=1$, the average chemical valance of Ni is Ni$^{2+}$, resulting in fully occupied $t_{2g}$ orbitals (not shown in Fig. \ref{fig1}) and two electrons remaining in the $e_{g}$ orbitals of each Ni atom. It is important to note that the strong bilayer coupling, further splitting two $d_{z^2}$ orbital into one bonding state $d_{z^2}^{+}$ and one anti-bonding state $d_{z^2}^{-}$, as shown in Fig. \ref{fig1}. However, the influence on $d_{x^2-y^2}$ orbitals can be neglected due to their geometric shape. In the bilayer block, there are a total of four electrons in the $e_{g}$ orbitals, and the $d_{x^2-y^2}$ orbitals are exactly half-filled.
This model is equivalent to a bilayer Hubbard model at half-filling \cite{bilayer_PhysRevB.73.245118,okamoto_PhysRevB.75.193103,Nb3Cl8_PhysRevB.107.035126}.

\begin{figure}
	\begin{center}
		\fig{3.4in}{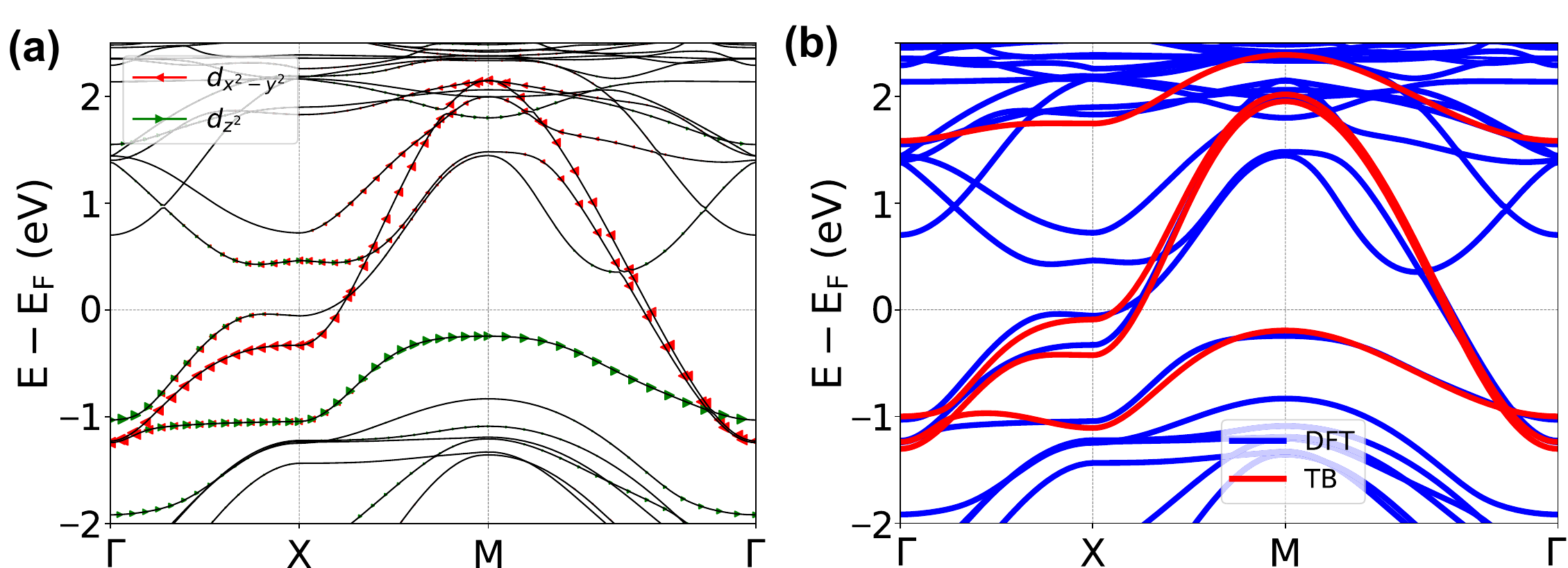}
		\caption{(a) The band structure from DFT calculation. The red and green dots represent the orbital projections of $d_{x^2-y^2}$ and $d_{z^2}$, respectively. (b) The comparison between the DFT calculation (blue lines) and bilayer two-orbital TB model (red lines). We can see the TB bands fit the $e_{g}$ bands very well around the Fermi level. 
			\label{fig2}}
	\end{center}
\end{figure}

We begin by considering $x=1$ and performing first-principle density functional theory (DFT) calculations to investigate the uncorrelated electronic structure. The oxygen vacancies are treated using the VCA discussed above. Our DFT calculations are carried out with the Vienna ab-initio simulation package (VASP) code \cite{kresse1996efficient} with the projector augmented wave (PAW) method \cite{kresse1999ultrasoft}. The generalized gradient approximation (GGA) exchange-correlation functional and its Perdew-Burke-Ernzerhof (PBE) version \cite{perdew1996generalized} is used. The cutoff energy for expanding the wave functions into a plane-wave basis is set to be 500 eV. The energy convergence criterion is 10$^{−8}$ eV. All calculations are conducted using the primitive cell to save time. The $\Gamma$-centered 9$\times$9$\times$9 k-meshes are used here. The calculated band structure is shown in Fig. \ref{fig2}(a). As seen, the $d_{x^2-y^2}$ bands are nearly half-filled, consistent with our previous crystal field analysis. In contrast to the GGA band structure of $x=0$, which have been extensively studied by various research groups, the bonding $d_{z^2}$ band here is fully occupied and inactive at Fermi level \cite{yxwang_PhysRevB.110.205122}. Additionally, the interlayer coupling is stronger at $x=1$ for the $d_{z^2}$ orbital, resulting in the anti-bonding $d_{z^2}$ band being located at a higher energy than $x=0$. 

Consistent with the previous study of low-energy model of conventional bilayer nickelate superconductors \cite{yxwang_PhysRevB.110.205122}, we have developed a tight-binding (TB) model for the $e_g$ orbitals of Ni atoms based on the results of DFT calculations \cite{mostofi2008wannier90,marzari2012maximally}. Since there are 2 atoms per unit cell, we derive a bilayer two-orbital (4-band) model $H_{TB}$. The dispersion of these TB bands is compared to the DFT bands in Fig. \ref{fig2}(b). As shown, the TB bands match the DFT bands very well around the Fermi level. Furthermore, we can extract the hopping parameters of this 4-band model. The corresponding Hamiltonian $H_{TB}$ and the explicit parameters can be found in Supplemental Materials (SM), which provides a faithful description of the low-energy electronic structure.

\begin{figure}
	\begin{center}
		\fig{3.4in}{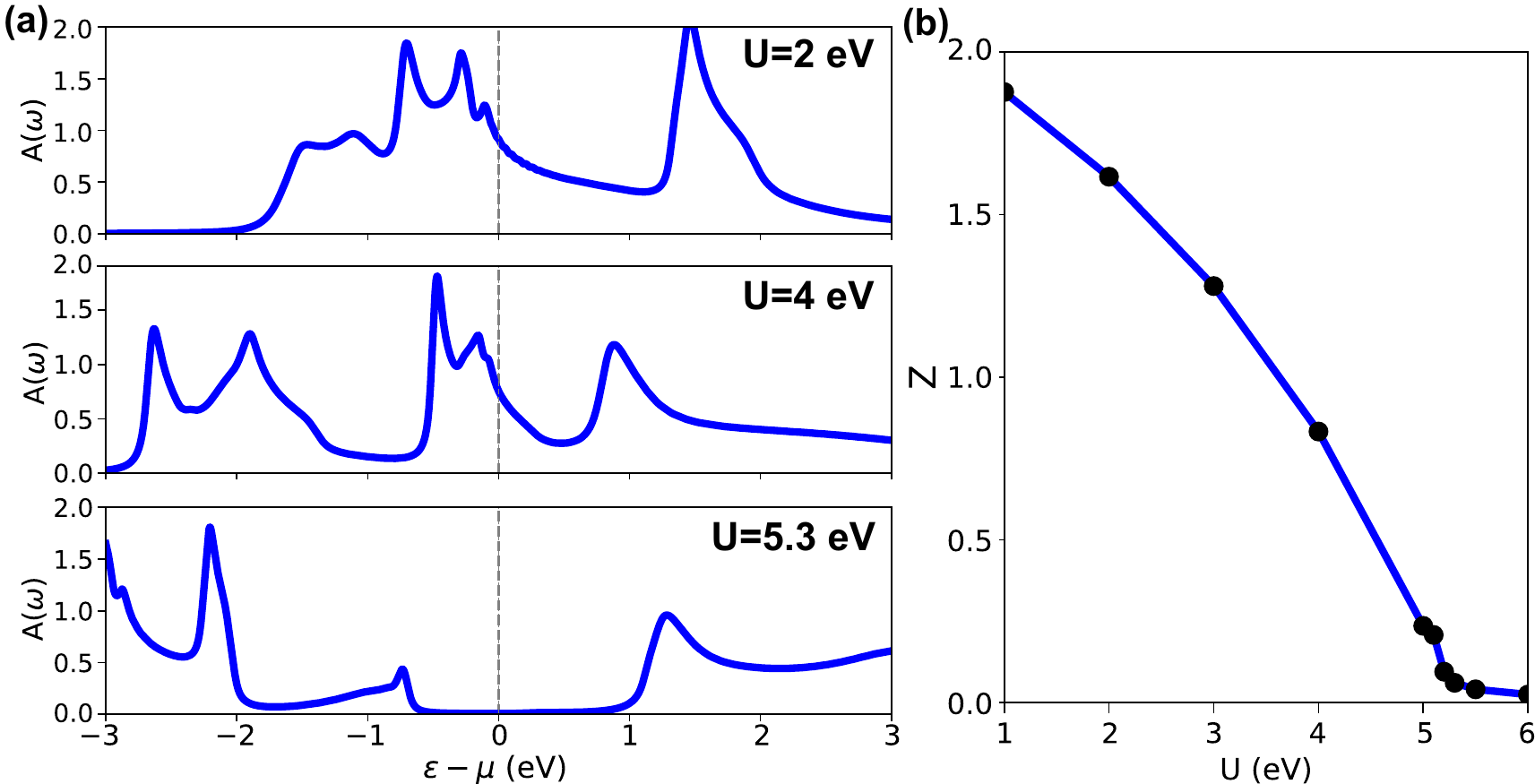}
		\caption{(a) The spectral functions $A(\omega)$ for $x=1$ obtained from DMFT calculation based on 4-band model are shown for $U=2$ eV, 4 eV and 5.3 eV, respectively. The $A(\omega)$ gradually evolves from a metallic state into a Mott insulator state, with the clear development of a Mott gap at $U=5.3$ eV. (b) The quasi-particle weight $Z$ at the Fermi level. The Mott transition occurs around $U=5.3$ eV, consistent with the figure (a).
			\label{fig3}}
	\end{center}
\end{figure}

The DFT band structure for $x=1$ clearly shows a metallic behavior, contradicting to the experimental findings \cite{ueki2024phase,cava_3265}. Therefore, it is necessary to include the correlation effect. For this bilayer two-orbital model, we consider the multi-orbital Hubbard interaction:
\begin{align}
	\begin{split}
		H_I&= U\sum_{l,i,\eta} \hat{n}_{l,i,\eta \uparrow}\hat{n}_{l,i,\eta \downarrow}+(U-2J_H)\sum_{l,i,\eta\neq \eta^{\prime}}\hat{n}_{l,i,\eta}\hat{n}_{l,i,\eta^{\prime}}
		\\
		& -J_H\sum_{l,i,\eta\ne \eta^{\prime}} (\mathbf{S}_{l,i,\eta} \cdot  \mathbf{S}_{l,i,\eta'} 
		+d_{l,i,\eta\uparrow}^{\dagger}d_{l,i,\eta\downarrow}^{\dagger}d_{l,i,\eta^{\prime}\uparrow}d_{l,i,\eta^{\prime}\downarrow} ),
	\end{split}
	\label{eq:HI}
\end{align}
with the $l=t,b$ denoting the layer index, $\eta$ the spin index and $i$ the site index. To obtain the correlated band structure for $x=1$, we perform the dynamical mean field theory (DMFT) calculations on the previously obtained TB results, adding this Hubbard interaction $H_I$. Our DMFT calculation is implemented using the open-source TRIQS package \cite{triqs} with the continuous-time quantum Monte Carlo (CT-QMC) impurity solver CTHYB \cite{ctqmc}. The analytical continuation is performed using the maximum entropy method implemented in Maxent package \cite{maxent}. To reduce the dimensionality of the parameter space, we fixed $J_{H} = 0.15U$. To determine the Mott transition point, we carry out DMFT calculations for various values of $U$ within the range of 1 eV to 6 eV. A criterion for identifying this transition point is that the quasi-particle weight $Z$ at the Fermi level approaches to zero. It is defined as
\begin{align}
    Z=\left[ 1-\frac{\partial \rm{Re}\Sigma(\omega)}{\partial \omega}\bigg| _{\omega=0} \right]^{-1},
\end{align}
where Re$\Sigma(\omega)$ is the real part of the electronic self-energy at real frequency. The dependence of $Z$ on $U$ is shown in Fig. \ref{fig3}(b). It is clear that $Z$ drops to zero at approximately $U=5.3$ eV. Therefore, we identify the Mott transition at this $U$ value.

To further confirm this, the spectral functions $A(\omega)$ at $U=2$ eV, $U=4$ eV and $U=5.3$ eV are also plotted in Fig. \ref{fig3}(a). We define the $W$ as the bandwidth, which is approximately 3-4 eV in this material (See Fig. \ref{fig2}(b)). At $U=2$ eV, the interaction is weak compared to the kinetic energy, and $A(\omega)$ can be regarded as the slightly modified non-interacting density of states. At $U=4$ eV, the interaction enters the intermediate regime with $U\approx W$. Thus, $A(\omega)$ should exhibit features from both the non-interacting limit $U\ll W$ and the atomic limit $U\gg W$. As a result, the characteristic three-peak structure commonly observed in strongly correlated materials emerges. As $U$ increases, the spectral weight gradually transfers from the central quasi-particle peak to the Hubbard bands on both sides. At the critical value of $U=5.3$ eV, the spectral clearly exhibits a Mott gap.

Hence, the Mott transition is identified for $x=1$ from the DMFT calculation. 
Experimentally, La$_3$Ni$_2$O$_{6.5}$ has been identified as a spin-singlet Mott insulator based on two observations. First, transport measurements show that La$_3$Ni$_2$O$_{6.5}$ is a large-gap insulator \cite{ueki2024phase,cava_3265}, with the resistivity exhibiting a clear Arrhenius-type temperature dependence—an established hallmark of Mott insulating behavior.
Second, magnetic susceptibility data yield a positive Curie–Weiss temperature of +7.5 K, which is inconsistent with long-range antiferromagnetic (AFM) Mott insulator that typically exhibits negative Curie–Weiss temperatures. For comparison, the canonical (S=1) AFM Mott insulator La$_2$NiO$_4$ shows a value of −500 K \cite{goodenough_la214,ganguly1984crystal}. This contrast is further supported by the small measured magnetic moment of Ni (0.75 $\mu_{B}$ from Curie–Weiss fitting) in La$_3$Ni$_2$O$_{6.5}$, significantly lower than the $\sim3\mu_{B}$ expected for Ni$^{2+}$ ($S=1$) ions in La$_2$NiO$_4$ \cite{goodenough_la214,ganguly1984crystal}.
Our result is consistent with the phase diagram findings of the bilayer single orbital Hubbard model in the non-magnetic case \cite{bilayer_PhysRevB.73.245118,okamoto_PhysRevB.75.193103,Nb3Cl8_PhysRevB.107.035126}. On the other hand, previous DFT+$U$ study \cite{Pickett} suggests that the band structure is insulating into an AFM state, which deviates from our results and experimental findings.

Before further discussion, we would like to emphasize that we employ the VCA approximation to construct our effective bilayer two-orbital model. While the VCA La$_3$Ni$_2$O$_{6.5}$ is not, strictly speaking, a realistic structure, its electronic properties are expected to closely approximate those of a more realistic system. X-ray diffraction results indicate that the symmetry group of La$_3$Ni$_2$O$_{6.5}$ aligns with $I4/mmm$, suggesting that oxygen vacancies are mostly likely randomly distributed \cite{cava_3265,ueki2024phase}. When considering the random distribution of oxygen vacancies, La$_3$Ni$_2$O$_{6.5}$ is likely to exhibit stronger insulating behavior. Consequently, our approach remains valid within this context.

Next, we want to address the Anderson transition from La$_3$Ni$_2$O$_7$ to La$_3$Ni$_2$O$_{6.5}$. 
We employ the dynamical cluster approximation (DCA) \cite{Jarrell_2001,Hettler_2000} and the typical medium dynamical cluster approximation (TMDCA) \cite{tmdca_review} to investigate localization effects. In both approaches, the original lattice model is mapped onto a finite-size cluster embedded in an effective medium. The effective medium is determined self-consistently through the cluster self-energy, allowing the method to capture the effect of both electronic correlations and disorder.

\begin{figure}
	\begin{center}
		\fig{3.4in}{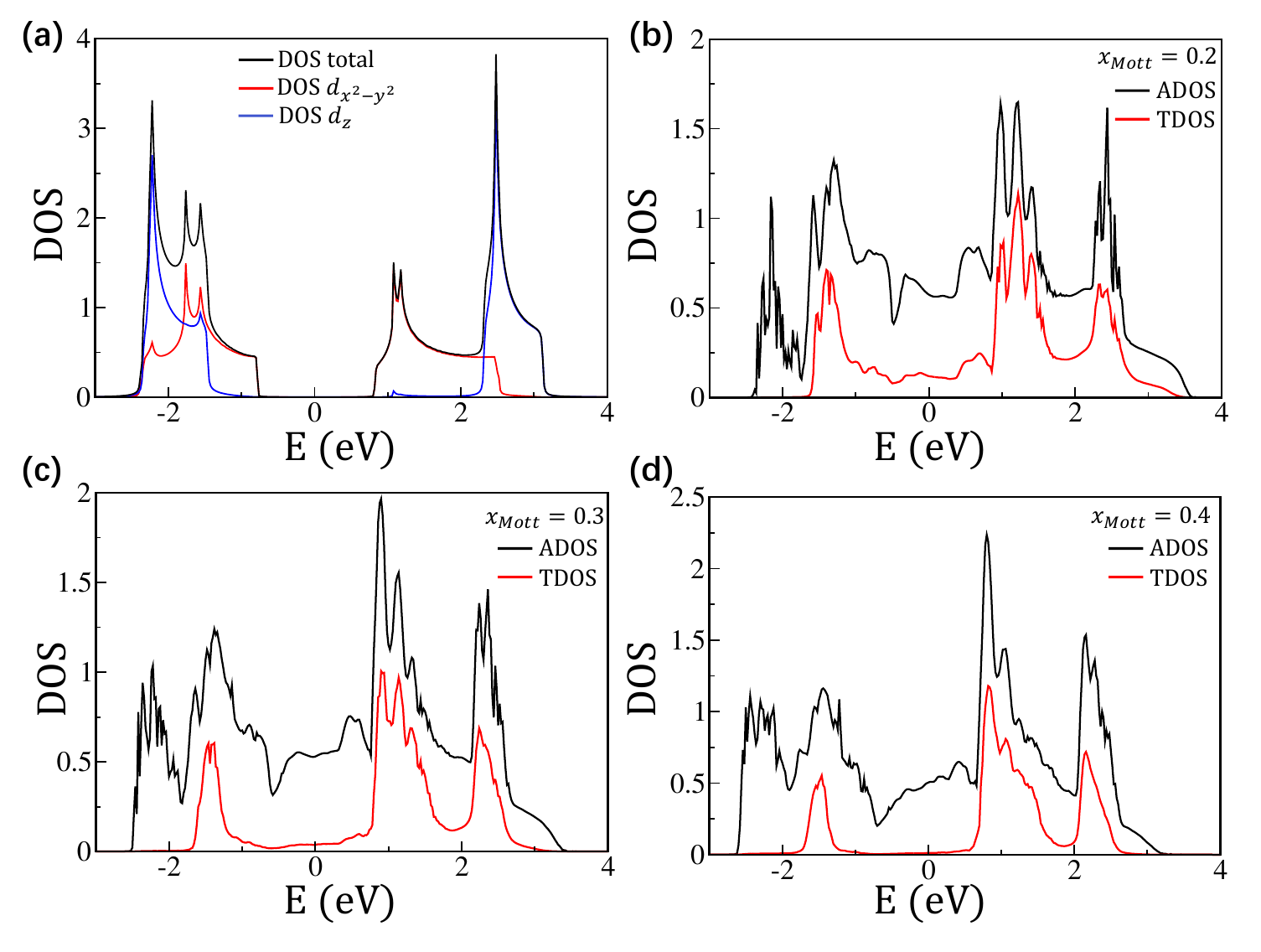}
		\caption{(a) The density of states of the pure Mott-insulating state obtained by introducing a Hubbard-I self-energy $\Sigma_H(\omega)$ to the La$_3$Ni$_2$O$_{6.5}$ bilayer two-orbital model. (b-d) Comparison of ADOS calculated by DCA and TDOS calculated by TMDCA with various concentration of the Mott states x$_{Mott}$ = 0.2, 0.3, 0.4. Both the DCA and TMDCA calculation use cluster size of $N_c=100$ and 2560 independent disordered realization. The Fermi level is determined by integrating the ADOS according to the doping induced by x$_{Mott}$ and is shifted to zero in all plots.
			\label{fig4}}
	\end{center}
\end{figure}

Specifically, we model the system as a binary mixture of the metallic state at $x = 0$ (La$_3$Ni$_2$O$_7$) and the Mott-insulating state at $x = 1$ (La$_3$Ni$_2$O$_{6.5}$).
The metallic state at $x=0$ can be described by a bilayer two-orbital TB model $H_{0}$, which has been studied in detail in Ref. \cite{yxwang_PhysRevB.110.205122}.
On the other hand, as discussed above, the state at $x = 1$ is well inside the Mott regime, which is driven by the correlation effect in the bilayer $d_{x^2-y^2}$ orbital and the bonding (antibonding) $d_{z^2}^{\pm}$ orbital is fully occupied (unoccupied).
To capture this physics, we employ the new-derived two-orbital bilayer TB model $H_{TB}$ for La$_3$Ni$_2$O$_{6.5}$ with a local Hubbard-I self-energy \cite{hubbard-I}, 
\begin{equation}
  \Sigma_{at}(\omega)=\frac{U_{eff}}{2} \frac{\omega+\mu_{eff}}{\omega+\mu_{eff}-\frac{U_{eff}}{2}-\mu_{eff}}  
\end{equation}
with $\mu_{eff}=\frac{U_{eff}}{2}$, acting only on the $d_{x^2-y^2}$ orbital to open the Mott gap. 
In addition, we introduce an effective interlayer coupling $t_p$ to account for the enhanced bonding–antibonding splitting of the $d_{z^2}^{\pm}$ orbitals in the insulating state.
Using parameters $U_{\mathrm{eff}} = 2.8$ eV and $t_p = -1.0$ eV, we obtain the total and orbital-resolved density of states (DOS) shown in Fig.~\ref{fig4}(a). 
The $d_{z^2}$ bands are well separated from the $d_{x^2-y^2}$. and the $d_{x^2-y^2}$ bands are split into upper and lower Hubbard bands with a Mott gap of about 1.5 eV.

The disorder-induced localization effect can be captured by comparing the average density of states (ADOS) and the typical density of states (TDOS) defined as the arithmetic and geometric average of the local density of states (LDOS), respectively. They can be expressed as
\begin{eqnarray}
    \text{ADOS}(E)=\left< \rho_i(E) \right>
    \\
    \text{TDOS}(E)=e^{ \left< \ln{\rho_i(E)} \right> }
\end{eqnarray}
where $\rho_i(E)$ is the total LDOS for each unit cell $i$ and $\left<\dots\right>$ corresponds to the average over different realizations of disorder and unit cell $i$. 
This is because the probability distributions of the LDOS for a weakly disordered metal and the strongly disordered Anderson insulator are very different. 
The distribution is Gaussian-like for the former and highly skewed as the log-normal distribution for the latter, where the TDOS approaches zero which can serve as an order parameter for the Anderson localization transition~\cite{tmt,tdos1}.
This approach is successfully applied to the single-band Anderson model~\cite{tmt,kpm,tdos1,tmdca} and the realistic systems~\cite{mbtmdca1,mbtmdca2} to study the localization effect.

We then apply DCA and TMDCA to the binary mixture to investigate Anderson localization. In our calculations, we use a cluster size of $N_c = 100$. For the cluster solver, we generate 2560 random disorder configurations, sampling each unit cell with probability $x_{\mathrm{Mott}}$ of being Mott insulating. This yields an effective oxygen vacancy concentration $\delta_{\mathrm{eff}} = 0.5x_{\mathrm{Mott}}$, corresponding to an effective doping $x_{\mathrm{eff}} = x_{\mathrm{Mott}}$.
Because both the local potentials and the nonlocal hopping parameters differ between the $x = 0$ and $x = 1$ models, the system exhibits both diagonal and off-diagonal disorder. We treat the off-diagonal disorder using the Blackman–Esterling–Berk (BEB) formalism \cite{beb_1,beb_2}, which has been successfully generalized to the TMDCA framework \cite{hanna_2014,mbtmdca2}.
After achieving self-consistency, we compute the ADOS within DCA and the TDOS within TMDCA.

The calculated ADOS and TDOS are shown in Figs.~\ref{fig4}(b-d) for three values of $x_{Mott}$. In the calculation, the Fermi level is determined by matching the integrated ADOS up to the Fermi level with the corresponding effective doping levels, which is shifted to zero in all the plots. 
When $x_{Mott}=0.2$, both ADOS and TDOS are finite at the Fermi level, which means the states around the Fermi level are still delocalized and the system is still metallic.
However, as $x_{Mott}$ increases, i.e. the oxygen vacancies increase, the TDOS at the Fermi level starts to decrease and approaches zero at $x_{Mott}=0.4$ while the ADOS remains finite at all values of $x_{Mott}$. This indicates that the states at the Fermi level become more and more localized as $x_{Mott}$ increases and the system becomes an insulator around the critical value at $x_c=0.4$ ($\delta_c=0.2$), where the TDOS at the Fermi level vanishes, signaling the Anderson localization transition, which is consistent with the experimental observations~\cite{ueki2024phase}.

In summary, we perform a theoretical investigation on the La$_3$Ni$_2$O$_{7-\delta}$ within the range of $\delta=0\sim0.5$.
For the $\delta=0.5$ phase, the low energy physics of La$_3$Ni$_2$O$_{6.5}$ is dominated by the bilayer $d_{x^2-y^2}$ orbital. 
Further considering the electron correlation effect in this bilayer model, a bilayer Mott insulator phase is identified through the DMFT calculation. This result matches well with recent experimental findings \cite{ueki2024phase,cava_3265}.
Furthermore, we consider the doping transition from $x=0$ to $x=1$ through mixing the metallic La$_3$Ni$_2$O$_{7}$ and insulating La$_3$Ni$_2$O$_{6.5}$. By evaluating the ADOS and TDOS at the Fermi level using DCA and TMDCA, we identify an Anderson localization transition driven by randomly distributed oxygen vacancies. This transition occurs at a critical doping of $x_c=0.4$ ($\delta_c=0.2$), in good agreement with recent experimental findings~\cite{ueki2024phase}.

More importantly, vacancy-induced insulating phases appear in close proximity to the superconducting dome of La$_3$Ni$_2$O$_{7}$ \cite{ueki2024phase}. In contrast, in the hole-doped regime—where disorder is not introduced through apical oxygen vacancies—both metallic behavior and superconductivity remain robust \cite{ueki2024phase}. These observations provide direct evidence that disorder-induced Anderson localization is a key mechanism underlying the suppression of superconductivity in La$_3$Ni$_2$O$_{7-\delta}$. This underscores the central role of oxygen stoichiometry and disorder in determining the phase diagram of nickelate superconductors.

We also notice that the La$_3$Ni$_2$O$_{7-\delta}$ thin film has been successfully synthesized \cite{erjia_guo,hwang,chen_zhuoyu}.
However, besides the compressive stress effects, the La$_3$Ni$_2$O$_{7-\delta}$ thin films always show insulating features at low-temperature \cite{erjia_guo,hwang,chen_zhuoyu}. Based on this work, we want to point out that the deficiency of oxygen content in these thin films may be one key reason for realizing the metallic phase and the superconducting phase \cite{hwang,chen_zhuoyu}. We hope our findings can provide a new perspective for insulating behaviors in La$_3$Ni$_2$O$_{7-\delta}$.

\begin{acknowledgments}
We acknowledge the support by the National Natural Science Foundation of China (Grant NSFC-12494594, No. NSFC-11888101, No. NSFC-12174428, and No. NSFC-12274279), the Ministry of Science and Technology  (Grant No. 2022YFA1403900), the New Cornerstone Investigator Program, and the Chinese Academy of Sciences Project for Young Scientists in Basic Research (2022YSBR-048). 
\end{acknowledgments}

\bibliography{ref}

\clearpage
\onecolumngrid
\begin{center}
\textbf{\large Supplemental Materials: The Mottness and the Anderson localization in bilayer nickelate La$_3$Ni$_2$O$_{7-\delta}$}
\end{center}

\setcounter{equation}{0}
\setcounter{figure}{0}
\setcounter{table}{0}
\setcounter{page}{1}
\makeatletter
\renewcommand{\theequation}{S\arabic{equation}}
\renewcommand{\thefigure}{S\arabic{figure}}
\renewcommand{\thetable}{S\arabic{table}}

\twocolumngrid

\section{The bilayer two-orbital tight-binding model for $x=1$ ($\text{La}_3\text{Ni}_2\text{O}_{6.5}$)}

In this section, we present the bilayer two-orbital TB model and parameters used in our calculations. We essentially adopt the bilayer two-orbital model from Ref. \cite{yxwang_PhysRevB.110.205122}. For a better fit, we have also included the third-nearest-neighbor intra-layer hopping $t_5$ between the same orbitals. The Hamiltonian, $H_{TB}(\textbf{k})$, is expressed in the basis $(d_{t\textbf{k}}^{x}, d_{t\textbf{k}}^{z}, d_{b\textbf{k}}^{x}, d_{b\textbf{k}}^{z})$ (with the spin index omitted) as:
\begin{align}
H_{TB} & ({\textbf{k}})=\left(\begin{array}{cc}
H_{t}({\textbf{k}}) & H_{\perp}({\textbf{k}})\\
H_{\perp}^{\dagger}({\textbf{k}}) & H_{b}({\textbf{k}})
\end{array}\right), \label{TB2}
\end{align}
where $H_{b}({\textbf{k}})=H_{t}({\textbf{k}})$.
Each block is a 2 $\times$ 2 matrix and is defined as follows:
\begin{align}
H_{t}({\textbf{k}})=\left(\begin{array}{cccc}
T_{{\textbf{k}}}^{x} & V_{{\textbf{k}}}\\
V_{{\textbf{k}}} & T_{{\textbf{k}}}^{z}
\end{array}\right), \label{ht}
\end{align}
and
\begin{align}
H_{\perp}({\textbf{k}})=\left(\begin{array}{cc}
t_{\bot}^{x} & V_{{\textbf{k}}}^{\prime}\\
V_{{\textbf{k}}}^{\prime} & t_{\bot}^{z}
\end{array}\right). 
\end{align}
The individual terms are defined as:
$T_{{\textbf{k}}}^{x/z}=t_{1}^{x/z}\gamma_k+t_{2}^{x/z}\alpha_k+t_{5}\delta_{k}+\epsilon^{x/z}$, $V_{\textbf{k}}=t_{3}^{xz}\beta_k$, $V_{\textbf{k}}^{\prime}=t_{4}^{xz}\beta_k$ with $\gamma_k=2(\cos k_x+\cos k_y)$, $\alpha_k=4\cos k_x\cos k_y$, $\beta_k=2(\cos k_x-\cos k_y)$ and $\delta_{k}=2(\cos 2k_x+\cos 2k_y)$. All parameters used in these expressions are summarized in Table. \ref{TB2}. 

\begin{table}[htbp!]
	\begin{tabular}{cccccc}
		\hline \hline 
		
		$t_{1}^{x}$ & $t_{1}^{z}$ & $t_{2}^{x}$ & $t_{2}^{z}$ & $t_{3}^{xz}$ & $t_{4}^{xz}$  \tabularnewline\hline 
		-0.4071 & -0.1007 & 0.0837 & 0.0403 & 0.1009 & -0.0022 \tabularnewline \hline 
		
	$t_{5}^{x}$ & $t_{5}^{z}$ &	$t_{\bot}^{x}$ & $t_{\bot}^{z}$ &  $\epsilon^{x}$ & $\epsilon^{z}$  \tabularnewline \hline 
		-0.0588 & -0.0187 & 0.0324 & -1.2914 & 0.258 & 0.6083 \tabularnewline
		
		\hline \hline
	\end{tabular}
\caption{The hopping parameters of the 4-band TB Hamiltonian $H_{TB}(\textbf{k})$ in unit of eV.} \label{TB2}
\end{table}

\section{The results of bilayer single-orbital tight-binding model and corresponding DMFT results for $x=1$ ($\text{La}_3\text{Ni}_2\text{O}_{6.5}$)}

\begin{figure}[htbp!]
	\begin{center}
		\fig{3.4in}{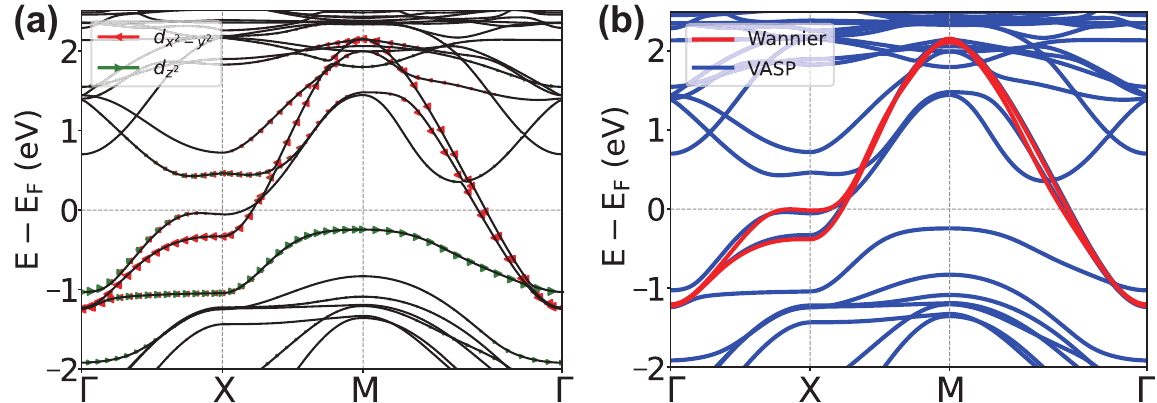}
		\caption{(a) The band structure from DFT calculation. (b) The comparison between the first-principles calculation and bilayer single-orbital model. We can see the wannier bands fit the $d_{x^2-y^2}$ bands very well around the Fermi level.
			\label{figS1}}
	\end{center}
\end{figure}

In fact, from the DFT bands, we can see that the $d_{z^2}$ orbital is inactive near the Fermi level. Therefore, an alternative simplified approach is to consider only the $d_{x^2-y^2}$ orbital. In this section, we present the results of the DMFT calculations using a bilayer single-orbital (2-band) model. The Wannier90 code \cite{mostofi2008wannier90,marzari2012maximally} is used to fit the $d_{x^2-y^2}$ bands obtained from previous DFT calculations. The dispersion of these wannierized bands is compared to the DFT bands in Fig. \ref{figS1}(b). As shown, the wannierized bands also match the DFT bands very well. 

The corresponding Hamiltonian $H_{TB-2}$ can be written as:
\begin{align}
H_{TB-2}({\textbf{k}})=\left(\begin{array}{cccc}
H_{11}(\textbf{k}) & H_{12}(\textbf{k}) \\
H_{21}(\textbf{k}) & H_{22}(\textbf{k})
\end{array}\right), 
\end{align}
where $H_{11}(\textbf{k})=H_{22}(\textbf{k})=\epsilon^{x}+t_{1}^{x}\gamma_{k}+t_{2}^{x}\alpha_{k}+t_{3}^{x}\delta_{k}$ and $H_{12}(\textbf{k})=H_{21}(\textbf{k})=t_{\perp}^{x}+t_{\perp1}^{x}\gamma_{k}+t_{\perp2}^{x}\alpha_{k}+t_{\perp3}^{x}\delta_{k}$ with $\gamma_{k}=2(\cos k_x+\cos k_y)$, $\alpha_{k}=4\cos k_x\cos k_y$ and $\delta_{k}=2(\cos 2k_x+\cos 2k_y)$. The corresponding on-site energies and hopping parameters are given in Table. \ref{TB}.

\begin{table}[htbp!]  
    \begin{tabular}{cccccccc}
		\hline \hline 
		$t_{1}^{x}$ & $t_{2}^{x}$ & $t_{3}^{x}$ & $t_{\perp}^{x}$ & $t_{\perp 1}^{x}$ & $t_{\perp 2}^{x}$ & $t_{\perp 3}^{x}$ & $\epsilon^{x}$  \tabularnewline\hline 
		-0.3902 & 0.0837 & -0.0333 & 0.0224 & 0.0089 
		& -0.0228 & 0.0219 & 0.2509 \tabularnewline
		\hline \hline
	\end{tabular}
    
	\caption{The hopping parameters of the 2-band TB Hamiltonian $H_{TB-2}(\textbf{k})$ in unit of eV.  $\epsilon^{x}$ is site energies for Ni $d_{x^2-y^2}$ orbital.  
	}
 \label{TB}
\end{table}

\begin{figure}[htbp!]
	\begin{center}
		\fig{3.4in}{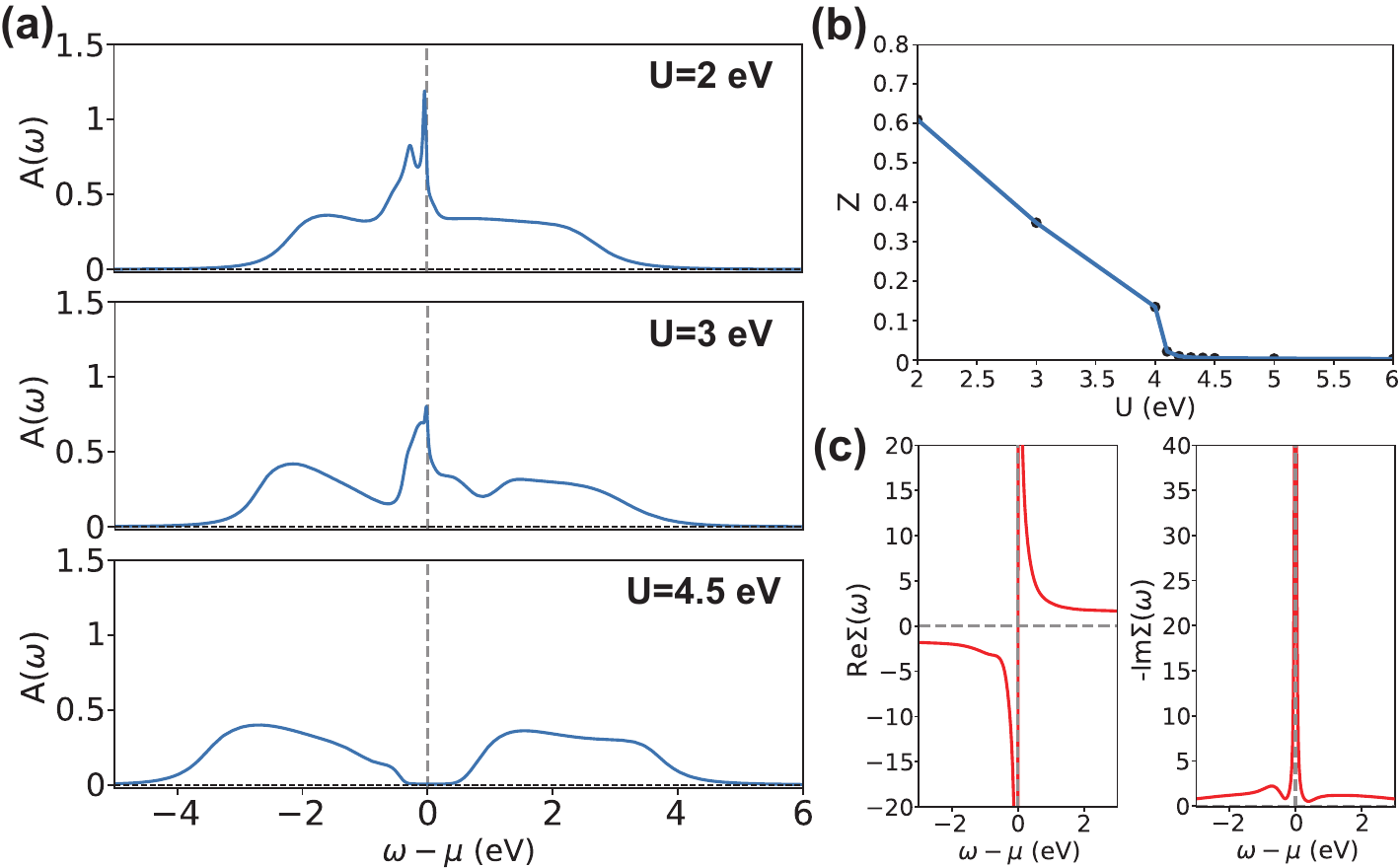}
		\caption{(a) The spectral functions $A(\omega)$ for $x=1$ obtained from DMFT calculation are shown for $U=2$ eV, $U=3$ eV and $U=4.5$ eV, respectively. The spectral evolves from the slightly modified non-interacting density of states to a three-peak structure, and ultimately to a clear Mott gap as the correlated strength $U$ increases. (b) The quasi-particle weight $Z$ at the Fermi level. The Mott transition occurs around $U=4.5$ eV. (c) The real (Re$\Sigma(\omega)$) and the negative imaginary parts (-Im$\Sigma(\omega)$) of the electronic self-energy at $U=4.5$ eV. Both of them exhibit the standard Mott insulating behavior.
			\label{figS2}}
	\end{center}
\end{figure}

Based on this model, we performed further DMFT calculations. We employed a single-orbital Hubbard interaction here:
\begin{align}
	\begin{split}
		H_I&= U\sum_{l,i} \hat{n}_{l,i \uparrow}\hat{n}_{l,i \downarrow},
	\end{split}
	\label{eq:HI-2}
\end{align}
where $l$ and $i$ denote the layer and site index, respectively.  For this case, we computed the spectral functions for different values of $U$ using DMFT, as shown in Fig. \ref{figS2}(a).

As shown in Fig. \ref{figS2}(b), the quasi-particle weight $Z$ at the Fermi level indicates a Mott transition occurring at around $U = 4.5$ eV. It can be seen that this value does not differ significantly from the result obtained in the bilayer two-orbital model (though that is evidently influenced by the choice of $J_{H}$). The corresponding electronic self-energy is displayed in Fig. \ref{figS2}(c). Both the real and imaginary parts demonstrate the typical behavior of a Mott insulator. Therefore, neglecting the contribution of the $d_{z^2}$ orbital is justified and does not qualitatively alter the main conclusions.

\section{The tight-binding model for $x=0$ ($\text{La}_3\text{Ni}_2\text{O}_{7}$)}

To investigate the Anderson transition in disordered systems, we employ the bilayer two-orbital model introduced in Ref. \cite{yxwang_PhysRevB.110.205122}, which describes the metallic state at $x=0$ (La$_3$Ni$_2$O$_7$). The Hamiltonian, $H_{0}(\textbf{k})$, is expressed in the basis $(d_{t\textbf{k}}^{x}, d_{t\textbf{k}}^{z}, d_{b\textbf{k}}^{x}, d_{b\textbf{k}}^{z})$ (with the spin index omitted), has exactly the same form as the Eq. \ref{TB2}, except that we are not considering third-neighbor coupling here, i.e., $t_5=0$. All parameters used in these expressions are summarized in the upper panel of Table~\ref{tab:Anderson}.

\begin{table}[htbp!]
	\begin{tabular}{ccccc}
		\hline \hline 
		
		$t_{1}^{x}$ & $t_{1}^{z}$ & $t_{2}^{x}$ & $t_{2}^{z}$ & $t_{3}^{xz}$  \tabularnewline\hline 
		-0.6003 & -0.149 & 0.0391 & -0.0007 & 0.2679 \tabularnewline \hline 
		
		$t_{\bot}^{x}$ & $t_{\bot}^{z}$ & $t_{4}^{xz}$ & $\epsilon^{x}$ & $\epsilon^{z}$  \tabularnewline \hline 
		0.038 & -0.999 & -0.072 &  1.2193 & 0.0048 \tabularnewline
		
		\hline \hline
	\end{tabular}
    
     \vspace{0.5cm}

    \begin{tabular}{ccccc}
		\hline \hline 
		
		$t_{1}^{x}$ & $t_{1}^{z}$ & $t_{2}^{x}$ & $t_{2}^{z}$ & $t_{3}^{xz}$  \tabularnewline\hline 
		-0.0153 & -0.0298 & 0.00782 & -0.00014 & 0.01 \tabularnewline \hline 
		
		$t_{\bot}^{x}$ & $t_{\bot}^{z}$ & $t_{4}^{xz}$ & $\epsilon^{x}$ & $\epsilon^{z}$  \tabularnewline \hline 
		-0.75 & -1.4985 & -0.0144 &  -0.148 & -0.1408 \tabularnewline
		
		\hline \hline
	\end{tabular}
    \caption{\label{tab:Anderson}The hopping parameters of the TB Hamiltonian used for the Anderson transition calculations. The upper panel corresponds to the bilayer two-orbital model $H_0$ describing the metallic state at $x$ = 0~\cite{yxwang_PhysRevB.110.205122}. The lower panel corresponds to $H_{Mott}$ used to describe the gapped band structure of the Mott insulator state at $x=1$. All parameters are in the unit of eV.    
	}
 \label{HP_TB}
\end{table}

\section{The model for the Mott state at $x=1$ ($\text{La}_3\text{Ni}_2\text{O}_{6.5}$)}

To describe the state at $x=1$ ($\text{La}_3\text{Ni}_2\text{O}_{6.5}$), we start with the bilayer two-orbital TB model $H_{TB}$ introduced above. To capture the Mott physics, we further introduce a local Hubbard-I self-energy
\begin{equation} 
  \Sigma_{at}(\omega)=\frac{U_{eff}}{2} \frac{\omega+\mu_{eff}}{\omega+\mu_{eff}-\frac{U_{eff}}{2}-\mu_{eff}}  
\end{equation}
with $\mu_{eff}=\frac{U_{eff}}{2}$, acting only on the $d_{x^2-y^2}$ orbital to open the Mott gap.
In addition, we introduce an effective interlayer coupling $t_p$ to account for the enhanced bonding–antibonding splitting of the $d_{z^2}^{\pm}$ orbitals in the insulating state.
The resulting self-energy matrix in the bilayer symmetrized basis is
\begin{equation}
    \Sigma_H(\omega)=\left(\begin{array}{cccc} 
		\Sigma_{at}(\omega) & 0 & 0 & 0\\
		   0 & t_p & 0 & 0\\
		   0 & 0 & \Sigma_{at}(\omega) & 0 \\
		   0 & 0 & 0 & -t_p 
	\end{array}\right) \ .
\end{equation}
The density of states shown in Fig. 4(a) of the main text is obtained from the imaginary part of the lattice Green’s function of the pure Mott-insulating state,
\begin{equation}
    G_H(k,\omega)=\frac{1}{\omega+i0^+-H_{TB}-\Sigma_{H}(\omega+i0^+)} \ ,
\end{equation}
summed over the Brillouin zone with chosen parameters as $U_{eff}=2.8$ eV and $t_p=-1$ eV.

\section{The Blackman–Esterling–Berk formalism used in DCA/TMDCA}

Since the model contains both diagonal and off-diagonal disorder, we treat the off-diagonal disorder using the Blackman–Esterling–Berk (BEB) formalism.
In particular, for the hopping parameters between two unit cells $i$ and $j$, we assume they depend only on the chemical occupations of the two sites. If we denote $i \in A$ for a metallic site and $i \in B$ for a Mott-insulating site, the hopping matrix elements are defined as
\begin{equation}
t_{ij,\alpha\beta}=\begin{cases}
\begin{array}{c}
(H_0)_{ij,\alpha\beta},\ \ \ \ \ \ \ \ \ if\ i \in A,\ j \in A\\
0,\ \ \ \ \ \ \ \ \ \ \ \ \ \ \ \ \ \ \ \ \ if\  i \in A,\ j \in B\\
0,\ \ \ \ \ \ \ \ \ \ \ \ \ \ \ \ \ \ \ \ \  if\  i \in B,\ j \in A\\
(H_{TB})_{ij,\alpha\beta}, \ \ \ \ \ \ \ if\ i \in B,\ j \in B,
\end{array}\end{cases}
\end{equation}
where $\alpha$ and $\beta$ label the layer and orbital indices.
This approximation effectively neglects hybridization between the metallic and Mott-insulating states.

\section{An alternative approach to study Anderson localization}

It is widely known that the bilayer Mott insulator smoothly connects with a bilayer band insulator \cite{bilayer_PhysRevB.73.245118,okamoto_PhysRevB.75.193103,Nb3Cl8_PhysRevB.107.035126}. Hence, we apply an alternative strategy of using a bilayer band insulator to mimic the bilayer Mott effect.
It turns out this is one efficient way to this Anderson localization problem. The tight-binding model, $H_{Mott}$, used to mimic the bilayer Mott effect is expressed in the same basis as $H_0$. The parameters for $H_{Mott}$ are listed in the lower panel of Table~\ref{tab:Anderson}. The impurity potential $V_{ij,\alpha\beta}=V_{ij,\alpha\beta}^{Mott}$ is just the difference between $H_{Mott}$ and $H_0$.

\begin{figure}
	\begin{center}
		\fig{3.4in}{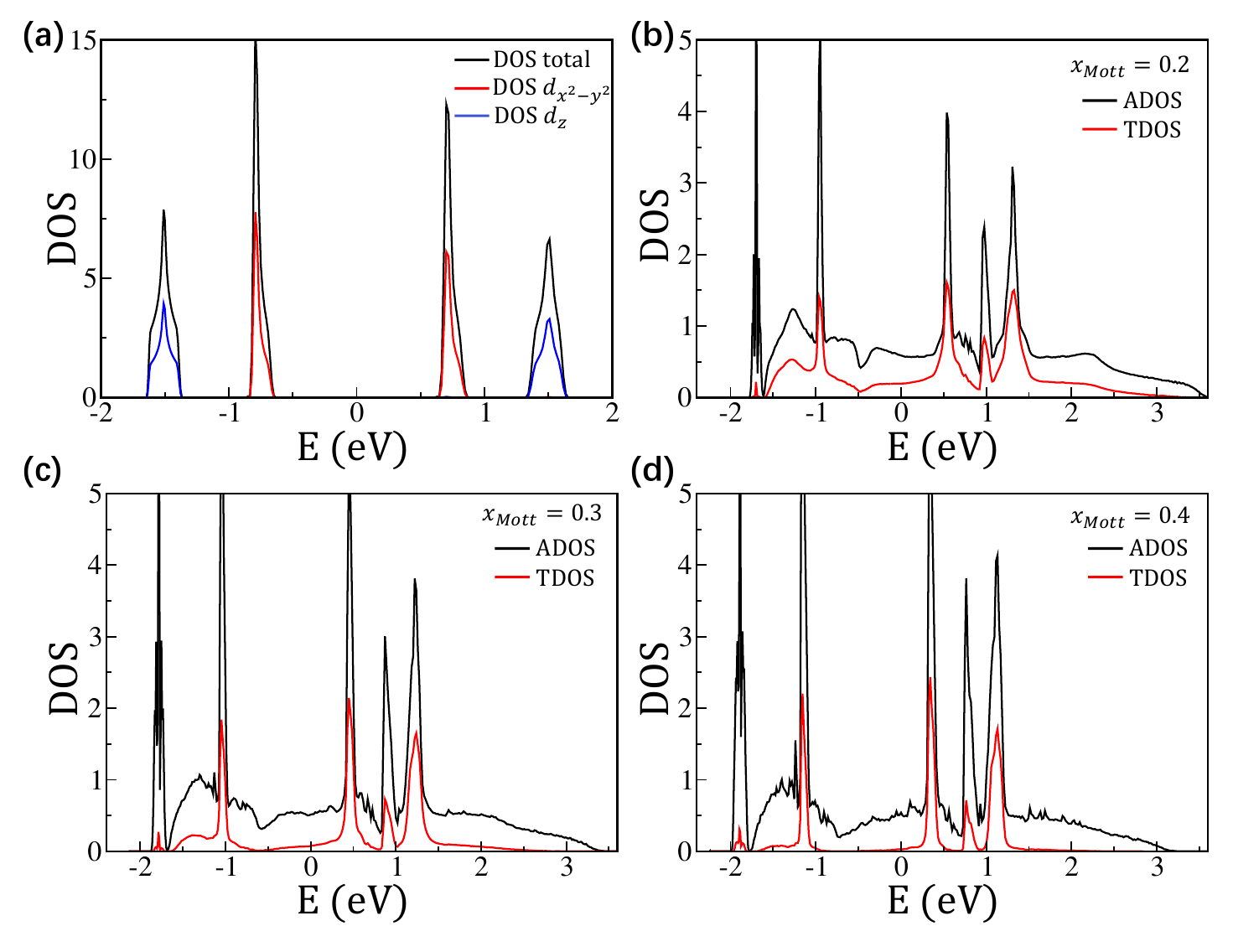}
		\caption{(a) The density of states of the band insulator used to simulate the Mott insulator where the total DOS is the sum of both layers, while the DOS of the two orbitals are for each layer. (b) Comparison of ADOS and TDOS calculated with KPM with various doping x = 0.2, 0.3, 0.4. The KPM uses 1024 moments on a square lattice with size 100$^2$ and 256 independent disordered realizations generated with 32 sites randomly sampled from each realization. The Fermi level is shifted to zero in all plots according to the doping induced by $x_{Mott}$.
			\label{fig_kpm}}
	\end{center}
\end{figure}

Next, we treat the system as a mixture of the metallic state at $x=0$ and the Mott insulator state at $x=1$ to study the disorder effect due to the randomly distributed oxygen vacancies.
Therefore, the Hamiltonian for this disordered system can be expressed as $H=H_0+H_{dis}$.
$H_{dis}$ describes the disorder effect and can be expressed as
\begin{equation}
    H_{dis}=\sum_{ij} \sum_{\alpha\beta} V_{ij,\alpha\beta} c_{i\alpha}^{\dagger}c_{j\beta}
\end{equation}
where $i, j$ are the indices of the unit cell and $\alpha, \beta$ are the combinations of the layer and orbital indices.
Here $V_{ij,\alpha\beta}=V_{ij,\alpha\beta}^{Mott}$ only when unit cells $i$ or $j$ are affected by the oxygen vacancy and become locally Mott-localized and vanishes otherwise.
$V_{ij,\alpha\beta}^{Mott}$ is determined by letting the Hamiltonian $H_{Mott}=H_{0}+\sum_{ij} \sum_{\alpha\beta} V_{ij,\alpha\beta}^{Mott} c_{i\alpha}^{\dagger}c_{j\beta}$ describe the gapped band structure of the Mott insulator at $x=1$. 
Then the real space Hamiltonian of the disordered system can be generated by considering each unit cell affected by the oxygen vacancy with the probability $x_{Mott}$, so that the effective oxygen vacancy becomes $\delta_{eff}=0.5x_{Mott}$, which leads to the effective doping $x_{eff}=x_{Mott}$.

As we discussed above, this Mott insulator is driven by the correlation effect in the bilayer $d_{x^2-y^2}$ orbital and the bonding (antibonding) $d_{z^2}^{\pm}$ orbital is fully occupied (unoccupied). Then, the effective bilayer model is constructed by increasing  $t_{\perp}^{x}$ and $t_{\perp}^{z}$. The total density of states (DOS) and the orbital projected DOS are plotted in Fig.~\ref{fig_kpm}(a). 
The $d_{z^2}$ bands are well separated from the $d_{x^2-y^2}$. and the $d_{x^2-y^2}$ bands are also split into two bands with gap of 1.3 eV, which is used to mimic the Mott gap for $x=1$.

We then calculate the ADOS and TDOS using the kernel polynomial method (KPM)~\cite{kpm,kpm2,kpm3}, where the LDOS is expanded by a series of Chebyshev polynomials whose arithmetic and geometric average can then be evaluated as the ADOS and TDOS.
The Jackson kernel is used in the KPM calculations~\cite{kpm}.
The calculated ADOS and TDOS are shown in Figs.~\ref{fig_kpm}(b-d) for three values of $x_{Mott}$. The evolution of the TDOS and ADOS shows similar behavior as that calculated from DCA and TMDCA in the main text, where critical concentration for the Anderson localization is around $x_c=0.4$ ($\delta_c=0.2$).
These results confirm that the simpler treatment of Mott insulator does not change the qualitative conclusions drawn from  including correlations explicitly within the DCA/TMDCA framework.

\end{document}